\documentclass{aastex}
\received{}
\accepted{}
\journalid{}{}
\articleid{}{}

\slugcomment{submitted to {\it Astrophysical Journal (Letters)}}

\begin{document}
\title
{An Intermittent Star Formation History in a `Normal' Disk Galaxy:
The Milky Way}

\author
{Helio J. Rocha-Pinto}
\affil{Instituto Astron\^omico e Geof\'{\i}sico da USP \\
        Av. Miguel Stefano 4200, 04301-904 S\~ao Paulo SP, Brazil}
\email{helio@iagusp.usp.br}
\author
{John Scalo}
\affil{University of Texas at Austin, USA}
\email{parrot@astro.as.utexas.edu}
\author
{Walter J. Maciel}
\affil{Instituto Astron\^omico e Geof\'{\i}sico da USP \\
        Av. Miguel Stefano 4200, 04301-904 S\~ao Paulo SP, Brazil}
\email{maciel@orion.iagusp.usp.br}
\and
\author
{Chris Flynn}
\affil{Tuorla Observatory, V\"aisa\"al\"antie 20, FIN-21500, Piikki\"o,
Finland}
\email{cflynn@astro.utu.fi}

\begin{abstract}
The star formation rate history of the Milky Way is derived using the
chromospheric age distribution for 552 stars in the solar neighborhood.
The stars' sample birthsites are distributed over a very large range of
distances because of orbital diffusion, and so give an estimate of the
global star formation rate history.  The derivation incorporates the
metallicity dependence of chromospheric emission at a given age, and
corrections to account for incompleteness, scale height--age correlations,
and stellar evolutionary effects.  We find fluctuations in the global star
formation
rate  with amplitudes greater than a factor of 2--3 on timescales less than
0.2--1 Gyr. The actual history is likely to be more bursty than found
here because of the smearing effect of age uncertainties.  There is
some evidence for a slow secular increase in the star formation rate,
perhaps a record of the accumulation history of our galaxy. A smooth
nearly-constant star
formation rate history is strongly ruled out, confirming the
result first discovered by Barry (1988) using a smaller sample and a
different age calibration. This result suggests that galaxies can fluctuate
coherently on large scales.
\end{abstract}


\section{Introduction}

The history of the average cosmic star formation rate (SFR) is of great
current interest, but is subject to severe uncertainties (see Pascarelle
et al. 1998, Glazebrook et al. 1998, Tresse \& Maddox 1998, Hughes et al.
1998, and
Cowie et al. 1999).  Such studies average over large numbers of galaxies, so
variations between galaxies and internal temporal variations within individual
galaxies are "washed out".  However, for an understanding of  the star
formation
process itself, and of individual galaxy evolution, it is just these
variations that
are of interest.

It is known that short-lived spatially coherent ``bursts" of star formation
occur on kiloparsec scales in starburst galaxies, giant H II
regions, ``superassociations"  (e.g. Efremov 1994), and in Local
Group dwarfs (see Tolstoy 1998,  Grebel 1997, Mateo 1998 and references
therein).  Since larger disk galaxies consist of spatially-connected
regions of gas of comparable size, and since propagation of star formation
is  well-established (see the comprehensive reviews by Elmegreen 1992,
1998),  it is unknown in the case of individual galaxies whether an entire
galaxy can  undergo some collective process that effectively synchronizes
global variations  in the SFR.

Some information on the SFR history of individual galaxies can be inferred
>from the ratio of present-to-past average SFR ratios in local galaxies
(see Kennicutt 1998 for a review), analysis of the color-magnitude diagrams
of Local Group dwarf galaxies (see Grebel 1997, Mateo 1998, Tolstoy 1998;
also Dolphin 1997,  Gallart et al. 1999), and recent pixel-by-pixel
modeling of the H$\alpha$ and UV luminosities
(Glazebrook et al. 1998).  But all of these studies attempt to model the
properties of whole populations of stars, which are subject to severe
assumptions
and uncertainties.  Clearly the most {\it direct} method for estimating the
SFR history of a galaxy is to use a determination of the ages of
individual stars in order to construct the age distribution.  The only sample
of relatively low-mass stars for which this approach can be used is the
sample of local stars in the Milky Way.
The estimation of this age distribution and the inferred SFR history is the
subject of the present {\it Letter}.

A crucial point is that the nearby
stars older than about 0.2 Gyr represent a large range in distances of their
birthsites.  Wielen (1977) showed that the orbital diffusion coefficient
deduced
>from the observed increase of velocity dispersion with age implies that
such stars have suffered an rms azimuthal drift of from about 2 kpc (for an
age of 0.2 Gyr)
to many galactic orbits (for an age of 10 Gyr). Considerable, but smaller,
drift should occur also in the radial  direction.  In this
sense the SFR inferred for nearby stars is a measure of the {\it global}
Milky Way SFR, at least at the Sun's galactocentric radius. More recent
estimates of the  diffusion coefficient (e.g. Meusinger et al. 1991) are
consistent with this conclusion.

Previous attempts to derive the age distribution of local stars have used
stellar evolutionary tracks (Twarog 1980, Meusinger 1991, Chereul, Crez\'e, \&
Bienaym\'e 1998), chromospheric  activity as measured by Ca II H and K emission
(Barry 1988, Soderblom, Duncan, \&  Johnson 1991), stellar kinematics
(G\'omez et
al. 1990, Marsakov et al.  1990), features in the main sequence (Scalo
1987) and
white dwarf (Noh \& Scalo 1990,  D\'{\i}az-Pinto et al. 1994, Isern et al.
1999) luminosity
functions,
combining the metallicity  distribution and age--metallicity relation of G
dwarfs
(Rocha-Pinto
\&  Maciel 1997), and the distribution of coronal
emission as measured by X-ray luminosities (Micela et al. 1993).  See also
Lachaume et al. (1999).  All these methods are fraught with difficulties.
However it  is notable that most of these studies
have inferred a SFR history that is non-monotonic with time.  There has
been a strong
tendency for astronomers to  overlook these results, partly because of the
lack of
appreciation of the importance  of orbit diffusion in making a local
stellar sample
representative of the global SFR  history, but also because a non-monotonic SFR
would provide unwanted complication  in galactic evolution studies and
provide a
foil to simple self-regulation models of  Galactic star formation, which
all yield a
smooth, monotonic SFR history.

The present {\it Letter} provides a new analysis of the SFR history based on
chromospheric  emission ages for a large sample of solar-like stars.  We
show that
it is very unlikely  that the Milky Way SFR history has been monotonic and
smooth,
and that it has  undergone fluctuations of at least a few (and probably much
larger).

\section{Chromospheric Ages}
The individual ages of the stars in our sample are based on
chromospheric emission.
The usual method of quantifying the observed chromospheric emission (CE) in
the Ca II H and K lines is based on the Mt. Wilson system of Vaughan et al.
(1978).
Corrections due to the fact that the continuum flux depends on the
photospheric
UV continuum, and due to photospheric light entering the instrumental
bandpasses,
yield a corrected quantity $R^\prime_{HK}$ as described in Noyes et al.
(1984).
A lower resolution estimate of $R^\prime_{HK}$, which could be
used to calibrate the CE-age relation using open clusters, was used in
Barry's (1988) estimate of the age distribution.  Soderblom et al. (1991) used
the higher-resolution system to calibrate the CE-age relation based on a
comparison of evolutionary tracks with Str\"omgren photometry of solar-type
stars that are secondaries in visual binaries, as
well as some slightly evolved F dwarfs, high velocity stars, the sun, and two
nearby clusters.  Rather than interpret the resulting age distribution as
non-monotonic, as found by Barry (1988), Soderblom et al. showed that a
nonlinear
CE-age relation, consistent with the available data, could yield a constant
SFR.

A major advance was the determination of $R^\prime_{HK}$ for a large number
of southern F-K (mostly G) dwarfs by Henry et al. (1996).  The present work
uses
this sample, supplemented by stars observed by Soderblom (1985).  The overlap
between this sample and stars that have published {\it uvby} photometry in
Olsen's
catalogues (Olsen 1983, 1993, 1994, needed to estimate the
metallicity-dependent
correction to the chromospheric ages found by Rocha-Pinto \& Maciel 1998)
yields 729 stars.
Hipparcos parallaxes are known for 714 of these stars.  The sample was reduced
by further considerations, mainly by omitting stars more distant than 80
pc, to
minimize any effect of reddening on
colors, and all stars with extremely strong CE ($\log R'_{\rm HK} \ge -4.20$),
which might be close binaries
instead of young stars (Soderblom et al. 1998).  The latter omission does
not affect our derived age distribution, since we are primarily concerned
with ages
greater than 0.1 Gyr and the number of stars omitted is small.  The final
sample
consists of 552 stars.

Given the $R^\prime_{HK}$ values from Henry et al. (1996) and Soderblom
(1985), ages
were calculated using the CE-age calibration given by eq. 3 of Soderblom et
al. (1991;
see also Donahue 1998).
This equation is a power law weighted fit to the $R^\prime_{HK}$ values and
ages of 42
stars and the sun, the Hyades cluster, and the Ursa Major group.  It will
be seen from our results that no reasonably smooth alteration of this
calibration
could eliminate the intermittent SFR history that we derive.  We emphasize,
however
that an improved CE-age calibration based on open clusters is sorely needed.

The chromospheric age of each star was corrected for metallicity dependence
with the relation derived
in Rocha-Pinto \& Maciel (1998), using the available {\it uvby} photometry
to estimate
metallicity.  The resulting age distribution was further corrected to
account for
the fact that the sample is not volume-limited, using a simple V/V$_{max}$
method to
assign a weight to each star according to the volume to which it could be
observed in a volume-limited sample.  We then corrected the age
distribution to
account for the fact that older stars have larger scale heights, since we want
to derive the SFR per unit area of the disk.  This correction used the
iterative procedure outlined in Noh \& Scalo (1990) using the average
scale height--mass relation given in Scalo (1986) and iterating on the mean
age
corresponding to each mass calculated from the observed age distribution.
Details will be presented elsewhere (Rocha-Pinto et al. 1999, hereafter
RPMSF).

Unresolved binaries present another source of uncertainty, which
depends in a complicated way on the distribution of mass ratios and the mass
of the primary.  For example, a G+K binary will appear younger than a single G
star because the chromospheric flux increases towards the redder stars, and
the combined flux of the pair will be larger than that presented by the G
dwarf alone.  Simulations of this effect, to be reported
elsewhere, indicate that the error in age is only 0.14\% for the stars
older than 3 Gyr,
and rises to 0.3\% for the youngest stars in the sample.  Overall, the
effect is
negligible compared to other sources of uncertainty for stars older than
about 0.5
Gyr.

The final transformation is from the age distribution to the SFR, which
involves stellar evolutionary effects.  Stars that have age $\tau$ are
those that
formed at time $T-\tau$ ago ($T$ = present age of the disk) and that have main
sequence lifetimes $\tau_{ms}>\tau$, so that they are still alive.  Then
the observed age distribution $g_{obs}(\tau)$ is related to the SFR history
$b(t)$ by
\begin{equation}
g_{abs}(\tau)=\int\limits^{\tau_{ms,max}}_\tau b(T-\tau)p(\tau_{ms})d\tau_{ms}
\end{equation}
where $p(\tau_{ms})$ is the probability distribution of main sequence
lifetimes of the sample and $\tau_{ms,max}$ is the maximum main sequence
lifetime
of stars in the sample, corresponding to the smallest mass ($\sim0.8\
M_\odot)$.
If $\tau$ is smaller than the minimum main sequence lifetime
$\tau_{ms,min}$ of
the stars in the sample, corresponding to the largest mass ($\sim1.4\
M_\odot$),
then all stars with these ages will be seen and no correction is required.
For
our sample, $\tau_{ms,min}$ is about 3 Gyr.

For ages larger than this, since $\tau_{ms}$ is a strong function of the
stellar mass, $p(\tau_{ms}$) is a transformation of the mass function of stars
in the sample.  The IMF is very uncertain in the 0.8-1.4 $M_\odot$ mass
region but
$p(\tau_{ms}$) is rather insensitive to the adopted IMF. We adopted the
Miller \& Scalo (1979) IMF, and have verified that the conclusions of this
{\it Letter} would not be affected by changes to the slope of the IMF power
law from
-1 to  -3.

The SFR history is then given by
\begin{equation}
b(t)=\frac{g_{obs}(\tau)}{\int_\tau^{\tau_{ms,max}}p(\tau_{ms})d\tau_{ms}}\ .
\end{equation}
for $\tau>\tau_{ms,min}$. The effect of the integral is to elevate the
observed
$g_{obs}(\tau)$ progressively for older stars.  An equivalent relation
between the
age distribution and the SFR history was  presented by Tinsley (1974).

\section{Results}

Figure 1a shows the raw age distribution for the sample as a histogram with
bins of width 0.2 Gyr uncorrected for metallicity effects.  In Fig. 1b the
effect of applying the metallicity-dependent age correction is shown.  Fig. 1c
shows the distribution, again including metallicity corrections, but with
weight
assigned to each star to correct for incompleteness based on $V/V_{max}$.
In Fig. 1d the iterative scale height correction has been applied to the
histogram of Fig. 1c; the effect is to progressively elevate the higher-age
bins relative to the lower-age bins, since the scale height increases with
age.
Note that the effect is not severe, and does not affect the general
structure of the fluctuations in the age distribution.

Figure 2 shows the SFR history (in units of the average SFR) obtained by
applying evolutionary corrections to the histogram of Fig. 1d.  The bin size
in Fig. 2 has been increased to 0.4 Gyr.  The error bars correspond to Poisson
counting uncertainties.  The figure shows fluctuations in the SFR of a
factor of at least a few: are
these fluctuations significant?
We have compared this SFR history with 6000 simulations of 552
stars each, drawn from a constant SFR.  The dotted
horizontal lines in Fig. 2  correspond to the 2$\sigma$ deviation expected
for a
constant SFR.  We have  compared the expected amplitudes of excursions from the
constant SFR case  simulations to the empirical result and find that the
probability
that the  empirical fluctuations are artifacts due to small number
statistics are
less than 2\% (details in RPMSF).
Considering that the
empirical fluctuations are correlated in time, the  probability that the
fluctuations are noise must be smaller than this estimate.  The derived
fluctuations in the SFR have a maximum value of about a
factor  of two to three.  However this is a lower limit because the age
uncertainties
effectively smear the age distribution. We speculate that the amplitude of the
resulting fluctuations may be an  order of magnitude, with timescales
significantly
smaller than shown in Fig. 2.

There is marginal evidence in Fig. 2 for a long term secular increase in
the SFR with time over many Gyr, perhaps consistent with the idea that our
galaxy has grown by the accumulation of smaller galaxies (see Unavase,
Wyse, and
Gilmore 1996 and references therein).  However this result
is tentative because the large timescale trend depends somewhat on the details
of our correction for scale height-age correlations and stellar evolutionary
effects.

\section{Discussion and Conclusions}

We have derived the SFR history of the Milky Way using chromospheric ages
for 552 stars in the solar neighborhood. The
results demonstrate rather conclusively that the SFR in our Galaxy has not
been monotonic with time, but instead exhibits significant fluctuations.
The details
of the form of the SFR history shown in Fig. 2 may be altered by changes in
the CE-age calibration, the metallicity correction, and other effects, so
that the exact times of ``bursts" and ``lulls" may be altered.  For example, a
comparison with times of close passage of the Magellanic Clouds (see RPMSF)

would be very uncertain.  However it does not seem possible to
us that
the finding of  significant fluctuations could be invalidated by such
effects.  For
example,  the application of the metallicity correction actually decreased the
amplitude of the fluctuations (see Fig. 1), and the corrections for scale
height
and evolution only introduce smooth, long timescale, modifications.   {\it
The SFR
history of the Milky Way has fluctuated on timescales less than  0.2--1 Gyr
with
amplitudes greater than a factor of 2--3}.  Thus we confirm  the result first
discovered by Barry (1988) based on a smaller sample and a  different CE-age
relation, although the form of the age distribution  found here
differs in
detail. The true SFR history has been smeared in our  derivation by substantial
uncertainties in the stellar ages, so the true SFR  history can only be
``spikier"
than derived here.

It is still conceivable that the irregularity in the derived SFR could be
an artifact
caused by a very nonlinear CE--age relation, as proposed by Soderblom et
al. (1991),
but the present sample is large enough that such a CE--age relation would
have to
be extremely irregular, and there is no observational or theoretical reason
to suspect such behaviour, while episodic galactic SFR histories are
well-known, at
least for smaller galaxies and starburts.

The disagreement between the ages of the oldest stars found here and the disk
age inferred from the dropoff  of the white dwarf luminosity function at small
luminosities (Winget et al. 1987; see
Knox, Hawkins, \& Hambly  1999 for an update) may be due to either an error
in the
white dwarf result or errors in the evolutionary tracks on which the CE-age
relation is based (Soderblom et al. 1991 used tracks from
Maeder 1976).  Most other methods for estimating the disk age only give lower
limits (e.g. Jimenez, Flynn, \& Kotoneva 1999), and so cannot be used to decide
between the two choices.  However, revisions in the age  calibration
derived from
evolutionary tracks should only contract (or expand) the time  axes in our
plots;
it seems impossible that such a revision could remove the  irregularity of
the SFR
that we have derived.

Finally, we note that our derived SFR history is qualitatively similar to
that derived by Glazebrook et al. (1998) for a sample of 13 field galaxies at
redshift about unity, using a generalization of the pixel-by-pixel population
synthesis method introduced by Abraham et al. (1998).  Glazebrook et al.
conclude that bursts dominate over the first $\sim$5 Gyr of the lives of their
sample galaxies, with intervals of 0.2--0.3 Gyr and durations 0.1--0.2 Gyr.
Pure
continuous SF is strongly ruled out, in agreement with our result for the
Milky Way.

\acknowledgments
This work was supported by FAPESP and CNPq to WJM and HJR-P, NASA Grant
NAG 5-3107 to JMS, and the Finnish Academy to CF. We thank Rob Kennicutt an
Ron Wilhelm for useful
comments, and
the referee for suggesting that we consider the effects of unresolved binaries.

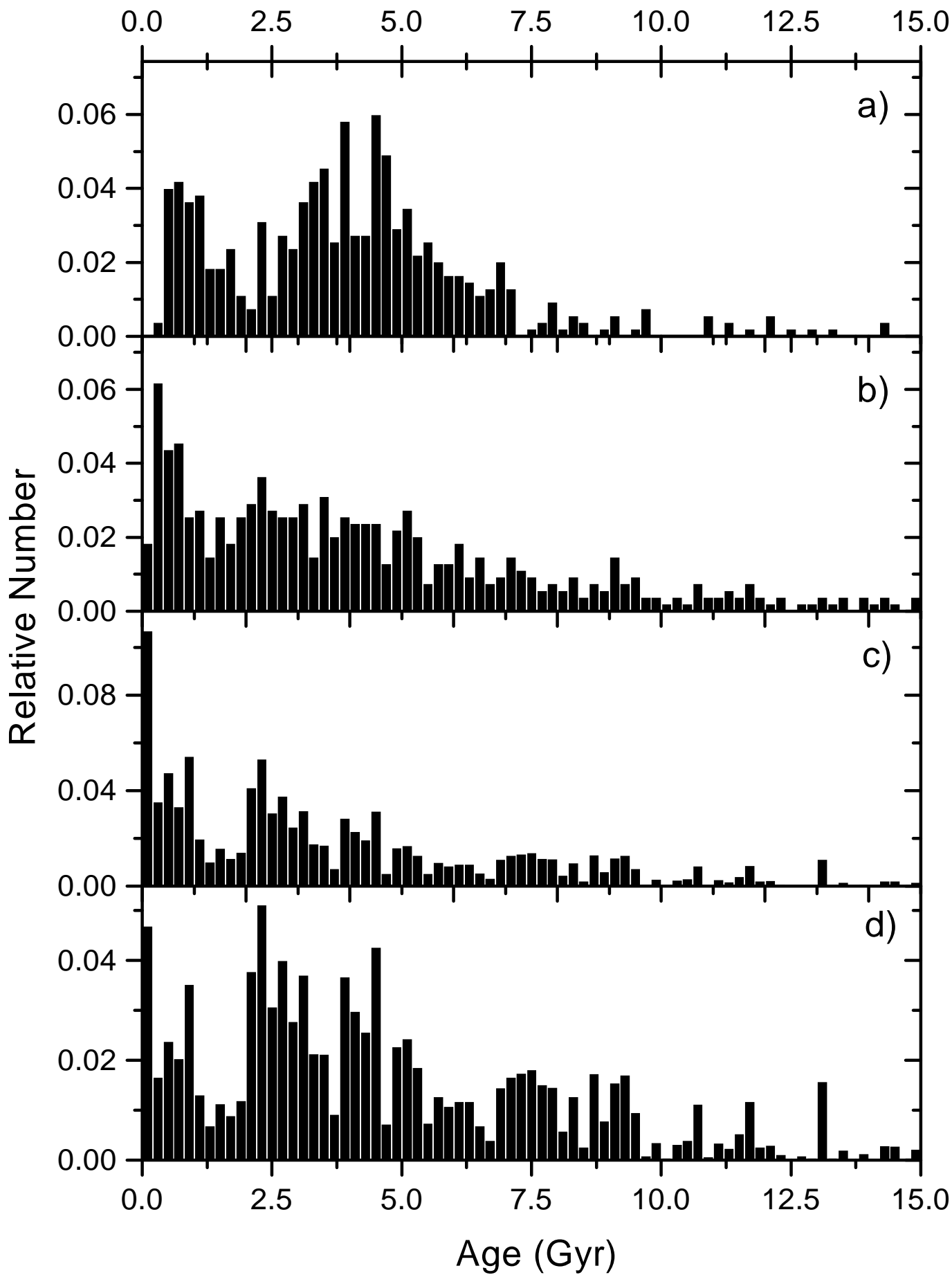
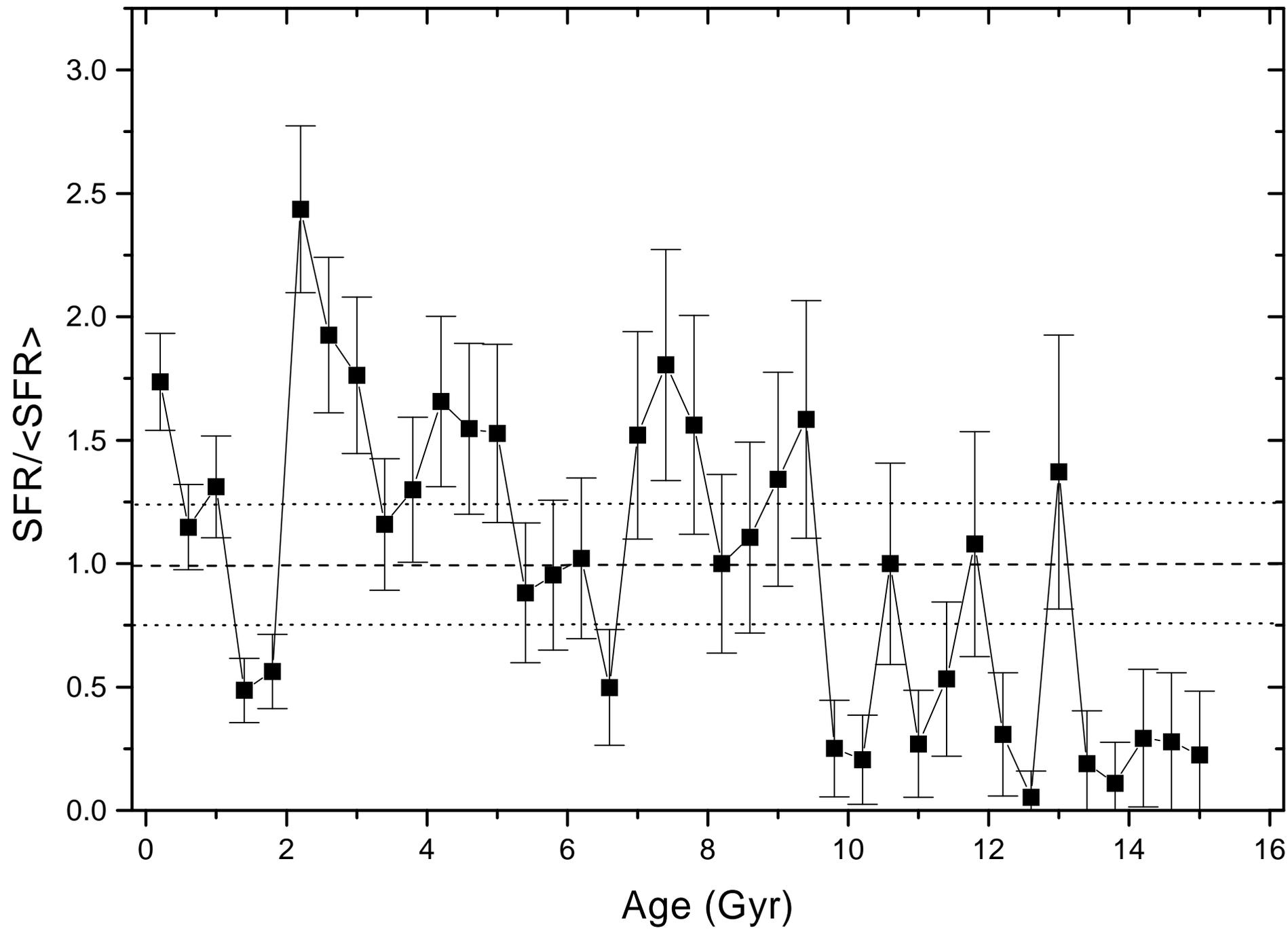
\end{document}